\documentclass[aoas,MSNbibl,nameyear,seceqn,dvips]{arximspdf}
\usepackage{multirow}
\usepackage{graphicx}

\doi{10.1214/14-AOAS719} 
\volume{8}
\issue{2}
\pubyear{2014}
\firstpage{999}
\lastpage{1021}

\makeatletter

\def\br{\mathbf{r}}
\def\bg{\mathbf{g}}
\def\bC{\mathbf{C}}
\def\bz{\mathbf{z}}
\def\bq{\mathbf{q}}
\def\ba{\mathbf{a}}
\def\bb{\mathbf{b}}
\def\bone{\mathbf{1}}
\def\bs{\mathbf{s}}

\def\bvarrho{\bolds{\varrho}}
\def\bomega{\bolds{\omega}}
\def\bpi{\bolds{\pi}}
\def\btheta{\bolds{\theta}}
\def\balpha{\bolds{\alpha}}

\def\mPr{\operatorname{Pr}}
\def\mN{\mathrm{N}}
\def\mIG{\operatorname{IG}}
\def\mDiscrete{\operatorname{Discrete}}
\def\mDirichlet{\operatorname{Dirichlet}}
\def\mG{\mathrm{G}}

\def\cDP{\mathcal{DP}}
\def\cA{\mathcal A}
\def\cP{\mathcal P}

\def\hmu{\widehat\mu}
\def\hsigma{\widehat\sigma}
\def\hp{\widehat{p}}

\def\wmu{\widetilde{\mu}}
\def\wsigma{\widetilde{\sigma}}
\def\wbtheta{\widetilde{\btheta}}
\def\wbphi{\widetilde{\bolds{\phi}}}
\def\wphi{\widetilde\phi}
\def\wq{\widetilde q}

\def\omu{\overline\mu}
\def\osigma{\overline\sigma}

\def\go{\rightarrow}
\newcommand{\eqref}[1]{(\ref{#1})}
\makeatother

\begin{document}
\begin{frontmatter}

\title{A Bayesian nonparametric mixture model for selecting
genes and gene subnetworks}
\runtitle{Bayesian gene and gene subnetwork selection}

\begin{aug}
\author{\fnms{Yize} \snm{Zhao}\thanksref{t1}\ead[label=e1]{yize.zhao@emory.edu}},
\author{\fnms{Jian} \snm{Kang}\corref{}\thanksref{t1,t2}\ead[label=e2]{jian.kang@emory.edu}}
\and
\author{\fnms{Tianwei} \snm{Yu}\thanksref{t3}\ead[label=e3]{tianwei.yu@emory.edu}}
\runauthor{Y. Zhao, J. Kang and T. Yu}
\affiliation{Emory University}
\thankstext{t1}{Equally contributed.}
\thankstext{t2}{Supported in part by the National
Center for Advancing Translational Sciences of
the National Institutes of Health under Award
Number UL1TR000454.}
\thankstext{t3}{Supported in part by the National Institutes of Health
grants P20 HL113451 and U19 AI090023.}
\address{Department of Biostatistics and\\
\quad Bioinformatics\\
Emory University \\
1518 Clifton Rd.\\ Atlanta, Georgia 30322\\USA\\
\printead{e1}\\
\hphantom{E-mail:\ }\printead*{e2}\\
\hphantom{E-mail:\ }\printead*{e3}} 
\end{aug}

\received{\smonth{5} \syear{2013}}
\revised{\smonth{11} \syear{2013}}

%
\begin{abstract}
It is very challenging to select informative features from
tens of thousands of measured features in high-throughput
data analysis. Recently, \mbox{several} parametric/regression models
have been developed utilizing the gene network information to
select genes or pathways strongly associated with a clinical/biological
outcome. Alternatively, in this paper, we propose a nonparametric
Bayesian model for gene selection incorporating network information.
In addition to identifying genes that have a strong association with a
clinical outcome, our model can select genes with particular expressional
behavior, in which case the regression models are not directly applicable.
We show that our proposed model is equivalent to an infinity mixture model
for which we develop a posterior computation algorithm based on Markov chain
Monte Carlo (MCMC) methods. We also propose two fast computing
algorithms that
approximate the posterior simulation with good accuracy but relatively low
computational cost. We illustrate our methods on simulation studies and the
analysis of Spellman yeast cell cycle microarray data.
\end{abstract}

%

\end{frontmatter}

\section{Introduction}
In high-throughput data analysis, selecting informative features from
tens of thousands of measured features is a difficult problem.
Incorporating pathway or network information into the analysis has been
a promising approach. Generally the setup of the problem contains two
pieces of information. The first is the measurements of the features in
multiple samples, typically with a clinical outcome associated with
each sample. The second piece of information is a network depicting the
biological relationship between the features, which is based on
existing biological knowledge. The network could contain information
such as protein interaction, transcriptional regulation, enzymatic
reaction and signal transduction, etc. [\citet{cerami2011pathway}].

Some methods are developed using the available network topology for
high-throughput data analysis. These methods incorporate the
gene-pathway relationships or gene network information into a
parametric/regression model. The primary goal is to identify either the
important pathways or the genes that are strongly associated with
clinical outcomes of interest. For example, there are a series of works
[\citeauthor{wei2007markov} (\citeyear{wei2007markov,wei2008markov});
\citeauthor{wei2010network} (\citeyear{wei2010network})] that model the gene
network using a Discrete or Gaussian Markov random field (DMRF or GMRF).
\citet{li2008network} and \citet{pan2010incorporating} used the gene
network to build penalties in a regression model for gene pathway
selection. \citet{ma2010incorporating} incorporated the gene
co-expression network in identification of caner prognosis markers
using a survival model. \citet{li2010bayesian} and \citet
{stingo2011incorporating} developed Bayesian linear regression models
using MRF priors or Ising priors that capture the dependent structure
of transcription factors or the gene network/pathway. Recently, \citet
{jacob2012more} proposed a powerful graph-structured two-sample test to
detect differentially expressed genes.

Although regression models are widely used for the selection of the
gene subnetwork that is associated with an outcome variable, in some
situations the question of interest is to study the expressional
behavior of genes, for example, periodicity, without an outcome variable.
In other situations, the experimental design is more complex than
simple case-control. For example, some gene expression studies involve
longitudinal/functional measurements for which the parametric models
[\citeauthor{leng2006classification} (\citeyear{leng2006classification});
\citeauthor{zhou2010analysis} (\citeyear{zhou2010analysis});
\citeauthor{breeze2011high} (\citeyear{breeze2011high})] or
the multivariate testing procedure [\citet{jacob2012more}] may not be
applicable without a major modification. A straightforward approach to
this problem is to perform large-scale simultaneous hypothesis testing
on gene behavior. A set of genes can be selected based on the testing
statistics or $p$-values, where a correct choice of a null distribution
for those correlated testing statistics [\citeauthor{efron2004large} (\citeyear{efron2004large,efron2010correlated})] should be used. However, this
approach ignores the gene network information that is useful to
identify the subnetwork of genes with the particular expressional
behavior. Due to the diverse behavior of neighboring genes on the
network, it is generally believed that genes in close proximity on a
network are likely to have joint effects on biological/medical outcomes
or have similar expressional behavior. This motivates the needs of
analyzing the large-scale testing statistics or statistical estimates
incorporating the network information. Another motivation is that a
linear regression or parametric model of gene expression levels might
not be suitable in some cases. For example, we may be interested in
finding subnetworks of genes that have nonlinear relations with an
outcome without specifying a parametric form. To address these
problems, a simple framework can be adopted. First, a certain statistic
is computed for each feature without considering the network structure.
The statistic can come from a test of nonlinear association, a test of
periodic behavior or a certain regression model. After obtaining the
feature-level statistics, a mixture model that takes into account the
network structure can be used to select interesting
features/subnetworks. More recently, \citet{qu2012hierarchical}
developed a Bayesian semiparametric model to take into account
dependencies across genes by extending a mixture model to a regression
model over the generated pseudo-covariates. This method could be
sensitive to the choices of the pseudo-covariates. \citet
{wei2012bayesian} proposed a Bayesian joint model of multiple gene
networks using a two-component Gaussian mixture model with a MRF prior.
This approach assumes the Gaussian distribution for each component
which might not fit the data very well in other applications.

To mitigate problems of the current methods, we propose a Bayesian
nonparametric mixture model for large-scale statistics incorporating
network information. Specifically, the gene specific statistics are
assumed to fall into two classes: ``unselected'' and ``selected,''
corresponding to whether the statistics are generated from a null
distribution, with prior probabilities $p_0$ and $p_1 = 1- p_0$. A
statistic has density either $f_0(r)$ or $f_1(r)$ depending on its
class, where $f_0(r)$ represents ``unselected'' density and $f_1(r)$
represents ``selected'' density. Thus, without knowing the classes, the
statistics follow a mixture distribution:
%
\begin{equation}\label{eq:lFDRmodel}
p_0 f_0(r) + p_1 f_1(r).
\end{equation}
As suggested by \citet{efron2010correlated}, it is reasonable to assume
statistics are normally distributed. This justifies the use of a
Dirichlet process mixture (DPM) of normal distributions to estimate
both $f_0(x)$ and $f_1(x)$. Note that different from \citet
{wei2012bayesian}, our model does not assume that $f_0$ and $f_1$
directly take the form of a normal density function. The DPM model has
been discussed extensively and widely used in Bayesian statistics
[\citeauthor{antoniak1974mixtures} (\citeyear{antoniak1974mixtures});
\citeauthor{escobar1994estimating} (\citeyear{escobar1994estimating});
\citeauthor{escobar1995bayesian} (\citeyear{escobar1995bayesian});
\citeauthor{muller2004nonparametric}
(\citeyear{muller2004nonparametric});\break
\citeauthor{dunson2010nonparametric} (\citeyear{dunson2010nonparametric})],
due to the availability of efficient computational techniques
[\citeauthor{neal2000markov} (\citeyear{neal2000markov});
\citeauthor{ishwaran2001gibbs} (\citeyear{ishwaran2001gibbs});
\citeauthor{wang2011fast} (\citeyear{wang2011fast})] and the
nonparametric nature with good performance on density estimation. The
DPM has been extended to make inference for differential gene
expression [\citet{do2005bayesian}] and estimate positive false
discovery rates [\citet{tang2007nonparametric}] but without
incorporating the network information. In our model, we assign an Ising
prior [\citet{li2010bayesian}] to class labels of all genes according to
the dependent structure of the network. As discussed previously, the
class label only takes two values, ``selected'' and ``unselected,'' while
a DPM model is equivalent to an infinity mixture model
[\citeauthor{neal2000markov} (\citeyear{neal2000markov});
\citeauthor{ishwaran2001gibbs} (\citeyear{ishwaran2001gibbs,ishwaran2002approximate})], based on
which we develop a posterior computation algorithm. Our method selects
genes and gene subnetworks automatically during the model fitting. To
reduce the computational cost, we propose two fast computation
algorithms that approximate the posterior distribution either using
finite mixture models or guided by a standard DPM model fitting, for
which we develop a hierarchical ordered distribution clustering (HODC)
algorithm. It essentially performs clustering on ordered density
functions. The fast computation algorithms can be tailored from any
routine algorithms for the standard DPM model and combined with the
HODC algorithm. Also, we suggest two approaches to choosing the
hyperparameters in the model.

To the best of our knowledge, our work is among the very first to
extend the DPM model to incorporate the gene network for gene
selections. Our method has the following features: (1) It provides a
general framework for gene selection based on large-scale statistics
using the network information. It can be used for detecting a
particular expressional/functional behavior, as well as the association
with a clinical/biological outcome. (2) It produces good uncertainty
estimates of gene selection from a Bayesian perspective, taking into
account the variability from many sources. (3) It introduces more
flexibility on model fitting adaptive to data in light of the
advantages of Bayesian nonparametric modeling. It is more robust than a
parametric model (e.g., two-component Gaussian mixture model) which is
sensitive to model assumptions. (4) The fast computational algorithms
have been developed for the posterior inference approximation. From our
experience, achieving a similar accuracy, it can be 50--150 times
(depending on the number of genes in the analysis) faster than the
standard Markov chain Monte Carlo (MCMC) algorithm. For a data set more
than 2000 genes, the analysis can be done within half an hour using a
typical personal computer. (5) Compared with the standard DPM model,
our model achieves much better selection accuracy in the simulation
studies and provides much more interpretable and biologically
meaningful results in the analysis of Spellman yeast cell cycle
microarray data. One potential issue of our method is that it only
allows the borrowing of information based on the network vicinity,
without considering possible compensatory effects between neighboring
genes. Such issues can be addressed by downstream analyses after a
small number of genes/subnetworks are selected.

The rest of paper is organized as follows. In Section~\ref{sec:model}
we describe the proposed model and an equivalent model representation.
We discuss the choice of priors and the details of the posterior
computation algorithms for gene selection. In Section~\ref{sec:fast} we
introduce fast computational algorithms for approximating posterior
computation. In Section~\ref{sec:app} we analyze an example data set, the Spellman
yeast cell cycle microarray data. We evaluate the performance of our
model via simulation studies in Section~\ref{sec: simulation}, where we
compare our results with a standard DPM model ignoring the network
information. We conclude the paper with discussions and future
directions in Section~\ref{sec:disc}.

\section{The model}\label{sec:model}
Let $n$ be the total number of genes in our analysis. For $i = 1,\ldots
, n$, let $r_i$ denote a statistic for gene $i$. It represents either a
functional behavior or the association with a clinical outcome. For the
association analysis, it is common to have an outcome $Y$ and a gene
expression profile $X_i$ for each gene, $i$. As an alternative to a
regression model, we can produce statistics for each gene, that is,
$r_i = s(X_i, Y)$, where $s(\cdot, \cdot)$ can be a covariance function
or other dependence test statistics. For a large-scale testing problem,
we usually obtain $p$-values, $p_1, \ldots, p_n$, which can be
transformed to normally distributed statistics, that is, $r_i = -\Phi
^{-1}(p_i)$, where $\Phi(r)$ denotes the cumulative distribution
function for the standard normal distribution. This transformation is a
monotone transformation and it ensures the ``selected'' genes have a
larger value of $r_i$.
Let $z_i$ be the class label for gene selection, where $z_i = 1$ if
gene $i$ is selected and $z_i = 0$ is unselected. For $i, j = 1,\ldots,
n$, let $c_{ij}$ denote the gene network configuration, where $c_{ij} =
1$ if gene $i$ and $j$ are connected, $c_{ij} = 0$ otherwise. Write
$\br= (r_1, \ldots, r_n)'$, $\bz= (z_1, \ldots, z_n)'$ and $\bC=
(c_{ij})$. In our model, $\br$ and $\bC$ are observed data, and $\bz$
is a latent vector of our primary interest.

\subsection{A network based DPM model for gene selection}
As suggested by \citet{efron2010correlated}, we assume $r_i$'s are
normally distributed. Let $\mN(\mu, \sigma^2)$ denote a normal
distribution with mean $\mu$ and standard deviation $\sigma$. Let
$\cDP
(G, \alpha)$ represent a Dirichlet process with base measure $G$ and
scalar precision $\alpha$. Given the class label $\bz$, we consider the
following DPM model: for $i = 1, 2, \ldots, n$ and $k = 0, 1$,
%
\begin{eqnarray}
\bigl[ r_i \mid\mu_i, \sigma^2_i
\bigr] &\sim& \mN\bigl(\mu_{i}, \sigma ^2_{i}
\bigr),\nonumber
\\
\label{eq:DPMmodel}\bigl[ \bigl(\mu_i, \sigma^2_i
\bigr)\mid z_i = k , G_k \bigr] &\sim& G_k,
\\
G_k &\sim& \cDP[G_{0k}, \tau_k],
\nonumber
\end{eqnarray}
where $\mu_i$ and $\sigma^2_i$ are latent mean and variance parameters
for each $r_i$. The random measure $G_k$ and the base measure $G_{0k}$
are both defined on $(-\infty, +\infty)\times(0, +\infty)$. We specify
$G_{0k} = \mN(\gamma_k, \xi^2_k) \times\mIG(\alpha_k, \beta_k)$, where
$\mIG(\alpha, \beta) $ denotes an inverse gamma distribution with shape
$\alpha$ and scale $\beta$. Note that
given latent parameters $\mu_i, \sigma^2_i$, the statistic $r_i$ is
conditionally independent of $z_i$. By integrating out $(\mu_i, \sigma
^2_i)$, we build the conditional density of $r_i$ given $z_i = k$ in
\eqref{eq:lFDRmodel}, that is,
%
\begin{equation}\label{eq:density}
f_k(r) = \int\pi(r\mid\btheta) \,d G_k (\btheta), \quad\quad\pi(r
\mid \btheta) = \frac{1}{\sigma}\phi \biggl(\frac{r-\mu}{\sigma} \biggr),
\end{equation}
where $\btheta= (\mu, \sigma^2)$ and $\phi(r)$ is the standard Gausian
density function. This provides a Bayesian nonparametric construction
of $f_k(r)$.

To incorporate the network structure, we assign a weighted Ising prior
to~$\bz$:
\begin{equation}\label{eq:zPrior}
\quad\pi(\bz\mid\bpi, \bvarrho,\bomega, \bC) \propto\exp \Biggl[ \sum
_{i=1}^n \biggl( \widetilde\omega_i
\log (\pi _{z_i}) + \varrho_{z_i} \sum
_{j \neq i} \omega_j c_{ij}
I[z_i=z_j] \biggr) \Biggr],
\end{equation}
where $\bpi= (\pi_0, \pi_1)$ with $0< \pi_1 = 1 - \pi_0 < 1$,
$\bvarrho= (\varrho_0, \varrho_1)$ with $\varrho_k > 0$ for $k = 0,1$,
$\bomega= (\omega_1, \ldots, \omega_n)'$ with $\omega_i>0$ for $i =
1,\ldots, n$, and $\widetilde\omega_i = \sum_{j=1}^n c_{ij} \omega
_j /\break
\sum_{j=1}^n c_{ij}$. The indicator function $I[\cA] = 1$ if event
$\cA
$ is true, $I[\cA] = 0$ otherwise.
The parameter $\bpi$ controls the sparsity of $\bz$, and the parameter
$\bvarrho$ characterizes the smoothness of $\bz$ over the network. For
each gene $i$, a weight $\omega_i$ is introduced to control the
information inflow to gene $i$ from other connected genes, which can
adjust the prior distribution of $z_i$ based on biologically meaningful
knowledge, if any. The term $\widetilde\omega_i$ is introduced to
balance the contribution from $\bpi$ and $\bvarrho$ to the prior
probability of $\bz$. When $\bvarrho=(0,0)$ and $\bomega= (1,\ldots
,1)'$, the latent class labels $z_{i}$'s are independent identically
distributed as Bernoulli with parameter $\pi_1$.

\subsection{Model representations}
As discussed by \citet{neal2000markov}, the DPM models can also be
obtained by taking the limit as the number of components goes to
infinity. With a similar fashion, we construct an equivalent model
representation of \eqref{eq:DPMmodel} for efficient posterior
computations. Let $\mDiscrete(\ba, \bb)$ denote a discrete distribution
taking values in $\ba= (a_1, \ldots, a_L)'$ with probability $\bb=
(b_1, \ldots, b_L)'$, that is, if $\xi\sim\mDiscrete(\ba, \bb)$, then
$\mPr(\xi= a_l) = b_l$, for $l = 1,\ldots, L$. Let $\mDirichlet
(\balpha
)$ denote a Dirichlet distribution with parameter $\balpha$. Let $L_k$,
for $k = 0, 1$, represent the number of components for density
$f_k(r)$. We define the index sets $\ba_0 = (-L_0+1, -L_0+2, \ldots,
0)$ and $\ba_1 = (1, 2,\ldots, L_1)$. Let $\bq_{0} =
(q_{-L_0+1},q_{-L_0+2},\ldots, q_{0})$ and $\bq_{1} = (q_{1},\ldots,
q_{L_1})$ with\vspace*{-3pt} $\sum_{g\in\ba_k} q_{g} = 1$. Let $\bone_{n} =
(\underbrace{1,\ldots,1}_n)$. Then model \eqref{eq:DPMmodel} is
equivalent to the following model, as $L_0\go\infty$ and $L_1 \go
\infty$:
%
\begin{eqnarray}\label{eq:finiteModel}
[ r_i \mid g_i, \wbtheta ] &\stackrel{\mathrm{i.i.d.}} {\sim}&
\mN\bigl(\wmu _{g_i},\wsigma^2_{g_i}\bigr),\nonumber
\\
 {[g_i \mid z_i = k, \bq_k ]}&
\stackrel{\mathrm{i.i.d.}} {\sim}& \mDiscrete (\ba _k, \bq_k),
\nonumber
\\[-8pt]\\[-8pt]
\wbtheta_{g} &\sim& G_{0k} \quad\quad \mbox{for } g\in
\ba_k,
\nonumber
\\
\bq_k &\sim& \mDirichlet(\tau_k\bone_{L_k}/L_k),
\nonumber
\end{eqnarray}
where $\wbtheta= \{\wbtheta_g\}_{g\in\ba_0 \cup\ba_1}$ and
$\wbtheta
_{g} = (\wmu_{g}, \wsigma^2_{g})$. The index $g_i$ indicates the latent
class associated with each data point $r_i$. Write $\bg= (g_1, \ldots,
g_n)$ and $\bz= (z_1,\ldots, z_n)$. For each class, $g$, the parameter
$\wbtheta_\mathrm{c}$ determines the distribution of $r_i$ from that class. The
conditional distributions of $g_i$ and $\wbtheta_{g_i}$ given $z_i = 0$
and $z_i = 1$ are different. Based on model \eqref{eq:finiteModel}, the
conditional density of $f_k(r)$ in \eqref{eq:density} becomes
%
\begin{equation}\label{eq:f_k}
f_k(r) = \sum_{g \in\ba_k} \frac{q_{g}}{\wsigma_g}
\phi \biggl(\frac
{r-\wmu_g}{\wsigma_g} \biggr).
\end{equation}
This further implies that given $L_0$ and $L_1$, the marginal
distribution of $r_i$ also has a form of finite mixture normals, that is,
%
\begin{equation}\label{eq:marginal}
\pi(r) = \sum_{k=0}^1 p_k
f_k(r) = \sum_{g=-L_0+1}^{L_1}
\frac{\wq
_g}{\wsigma_g}\phi \biggl(\frac{r-\wmu_g}{\wsigma_g} \biggr),
\end{equation}
where $\wq_g = p_0q_g$ if $g\leq0$, $\wq_g = p_1q_g$ otherwise.

Model \eqref{eq:finiteModel} is not identifiable for $z_i$ in the sense
that if we switch the gene selection class label ``$0$'' and ``$1$,'' the
marginal distribution of $r_i$ \eqref{eq:marginal} is unchanged.
Without loss of generality, we assume that the ``selected gene'' should
be more likely to have large statistics compared to the ``unselected
genes.'' Thus, we impose an order restriction on the parameter $\wbtheta
$, for $g = -L_0+1 ,\ldots, L_1$,
%
\begin{equation}
\wmu_g < \wmu_{g+1}.
\end{equation}
This also sorts out the nonidentifiability of parameter $\wbtheta$. In
many cases, the functional behaviors of some genes are strongly evident
from prior biological knowledge. Whether or not those genes are
selected is not necessarily determined by other genes in the network.
Those genes are likely to be the hubs of the networks, thus, the
determination of the status of these genes might help select genes in
their neighborhood. This suggests that it is reasonable to preselect a
small amount of genes that can be surely elicited by biologists from
their experience and knowledge. We refer to them as ``surely selected''
(SS) genes. These genes are usually associated with very large
statistics. We evaluate the performance via the simulation studies in
Section~\ref{subsec:s2}.

\subsection{Posterior computation}

In model \eqref{eq:finiteModel}, given $L_0$ and $L_1$, we have the
full conditional distribution of $g_i = g$ and $z_i = k$ given $\bg
_{-i} = (g_1, \ldots, g_{i-1},\break  g_{i+1},\ldots, g_{n})$, $\bz_{-i} =
(z_1, \ldots, z_{i-1}, z_{i+1}, \ldots, z_n)$ and data $\br$:
%
\begin{eqnarray}\label{eq:fullconditional}
&&\pi(g_i = g, z_i = k \mid\bg_{-i},
\bz_{-i}, \br, \wbtheta ) \nonumber
\\
&&\quad\quad\propto\frac{1}{\wsigma_g} \phi \biggl(\frac{r_i - \wmu
_g}{\wsigma
_g} \biggr)
\frac{n_{-ig}+\tau_k/L_k}{\tau_k + m_k -1}
\\
&&\quad\quad\quad{}\times\exp \biggl( \widetilde\omega_i \log(
\pi_{k}) + \varrho_{k} \sum_{j \neq i}
\omega_j c_{ij} I[z_j = k] \biggr),
\nonumber
\end{eqnarray}
where $m_k = \sum_{i=1}^nI[z_i=k]$ is the number of genes in class $k$
and $n_{-ig} = \sum_{j \neq i} I[g_{j} = g]$ represents the number of
$g_{j}$ for $j\neq i$ that are equal to $g$.

As $L_0 \go\infty$ and $L_1\go\infty$, if $(g, k) = (g_{j}, z_{j})$ for
some $j\neq i$, then
%
\begin{eqnarray}\label{eq:g_i=g}
&&\pi(g_i = g, z_i = k \mid\bg_{-i},
\bz_{-i}, \br, \wbtheta) \nonumber
\\
&&\quad\quad\propto\frac{n_{-ig}}{\tau_k + m_k -1} \exp \biggl( \widetilde \omega_i \log(
\pi_{k}) + \varrho_{k} \sum_{j \neq i}
\omega_j c_{ij} I[z_j = k] \biggr)
\\
&&\quad\quad\quad{}\times
\frac{1}{\wsigma_g} \phi \biggl(\frac{r_i - \wmu
_g}{\wsigma
_g} \biggr)
\nonumber
\end{eqnarray}
and
%
\begin{eqnarray} \label{eq:g_ineqg}
&&\pi(g_i \neq g_j, z_i \neq
z_j \mbox{ for all } j\neq i \mid \bg_{-i},
\bz_{-i}, \br, \wbtheta) \nonumber
\\
&&\quad\quad\propto\frac{\tau_k}{\tau_k + m_k -1} \exp \biggl( \widetilde \omega_i \log(
\pi_{k}) + \varrho_{k} \sum_{j \neq i}
\omega_j c_{ij} I[z_j = k] \biggr)
\\
&&\quad\quad\quad{} \times\frac{\Gamma(\alpha_k+1/2)\beta_k^{\alpha
_k}}{{\sqrt {2\pi}\Gamma(\alpha_k)\xi_k}} \int\phi \biggl(\frac{\mu-\gamma
_k }{\xi
_k} \biggr) \biggl(
\beta_k+\frac{1}{2}(r_i -\mu)^2
\biggr)^{-(\alpha
_k+1/2)} \,d \mu,
\nonumber
\end{eqnarray}
where the integral can be efficiently computed by the Gaussian
quadrature method in practice. See Section A in the supplemental
article [\citet{zhao2014bayesian}] for the derivations of equations
\eqref{eq:g_i=g} and \eqref{eq:g_ineqg}.

The full conditionals of $\wmu_g$ and $\wsigma^2_g$ for $g\in\{g_1,
\ldots, g_n \}$ are given by
%
\begin{eqnarray}
 \bigl[\wmu_g \mid\wsigma^2_g, \br
\bigr]&\sim&\mN \biggl(\frac{\wsigma
^2_g\gamma_k + \xi^2_k\sum_{i:g_i = g}r_i}{\wsigma^2_g+\xi_k^2n_g}, \frac{\wsigma^2_g\xi^2_k}{\wsigma^2_g+\xi^2_k n_g} \biggr),
\label
{eq:cond_mu}
\\
 \bigl[\wsigma^2_g \mid\wmu_g, \br
\bigr] &\sim&\mIG \biggl(\alpha_k+ \frac
{n_g}{2},
\beta_k + \frac{1}{2}\sum_{i: g_i = g}
(r_i - \wmu_g)^2 \biggr), \label{eq:cond_sigma}
\end{eqnarray}
where $k = I[g>0]$ and $n_g = \sum_{i=1}^n I[g_i = g]$. We summarize
this algorithm in Section B.1 in the supplemental article [\citet
{zhao2014bayesian}] and refer to it as NET-DPM-1. It is computationally
intensive when $n$ is very large. To mitigate this problem, we propose
two fast algorithms to fit finite mixture models (FMM) with appropriate
choices of the number of components.

\subsection{Fast computation algorithms}\label{sec:fast}
\subsubsection{FMM approximation}
When $L_1$ and $L_0$ fit the data well, we can accurately approximate
the infinite mixture model \eqref{eq:DPMmodel} by the FMM \eqref
{eq:finiteModel}. Given a fixed $L_0$ and $L_1$, it is straightforward
to perform posterior computation for model \eqref{eq:finiteModel} based
on \eqref{eq:fullconditional}. We refer to this algorithm as NET-DPM-2
(see Section B.2 in the supplemental article [\citet{zhao2014bayesian}]
for details). This algorithm does not change the dimension of $\wbtheta
$ over iterations. In this sense, it simplifies the computation. Also,
in order to keep computation efficient, we search for smaller values of
$L_0$ and $L_1$ which fit the data well. This can be achieved under the
guidance of a DPM density fitting for which we introduce an algorithm
in the next section.




\subsubsection{Hierarchical ordered density clustering}
Without using the network information, a DPM model fitting on data $\br
$ provides an approximation to the marginal density \eqref
{eq:marginal}. It generates posterior samples for mixture densities,
where the mean number of components should be close to $L_0+L_1$. Let
us focus on one sample. Suppose $L_0+L_1$ is equal to the number of
components in this sample. To further obtain an estimate of $L_0$ and
$L_1$ for this sample, we need to partition the $L_0+L_1$ components
into two classes. Thus, we propose an algorithm to cluster a set of
ordered densities. We call it hierarchical ordered density clustering
(HODC). Here, the density order is determined by the mean location of
that density. For example, a set of Gaussian density functions are
sorted according to their mean parameters.
Similar to the classical hierarchical clustering analysis, we define a
distance metric of density functions:
%
\begin{equation}
d\bigl(f, f'\bigr) = \int_{-\infty}^{+\infty}
\bigl[f(x) - f'(x)\bigr]^2 \,d x,
\end{equation}
where $f$ and $f'$ are two univariate density functions. Let $\cP= \{
(\hmu_g, \hsigma^2_g, \hp_g)\}_{g=1}^{L_0+L_1}$ denote parameters for
$L_0+L_1$ Gaussian densities, where $\hmu_g < \hmu_{g+1}$, $g=1,2,\dots
,L_0+L_1-1$. This is the input data to the HODC algorithm totally
consisting of $L_0+L_1 - 2$ steps. At the $m$ step, there are
$L_0+L_1-m$ clusters of densities and let $\bs^{(m)}_l$, for $l =
1,\ldots, L_0+L_1-m$, denote the density indices in cluster $l$.
To simplify, we define
%
\begin{equation}\label{eq:wphi}
\wphi(r; \bs,\cP) =\sum_{g\in\bs} \frac{\hp_{g}}{\hsigma_g}
\phi \biggl(\frac{r - \hmu_g}{\hsigma_g} \biggr) \Big/\sum_{g \in\bs}\hp
_{g},
\end{equation}
which represents a mixture of Gaussian densities, where the components
indexed by $\bs$ are a subset of $\{\phi[(r-\hat\mu_g)/\hat\sigma
_g]/\hsigma_g\}_{g = 1}^{L_0+L_1}$.\vadjust{\goodbreak}



\noindent\textbf{\underline{HODC}:}
\begin{itemize}
\item[]{\bf Input:} Parameters for a mixture of Gaussian densities,
that is, $\cP$.
\vspace{0.2cm}
\item[]{\bf Initialization:} Set $m = 0$ and $\bs_l^{(0)} = \{l\}$, for
$l = 1,2,\ldots, L_0+L_1$;
\vspace{0.2cm}
\item[] Repeat the following steps until $m =L_0+L_1-2$:
\begin{itemize}
\item[]{\bf Step 1:} Find
\[
l^{(m)} = \arg\min_{l} d \bigl(\wphi\bigl(\cdot;
\bs^{(m)}_{l},\cP\bigr), \wphi \bigl(\cdot;
\bs^{(m)}_{l+1},\cP\bigr) \bigr).
\]
\item[]{\bf Step 2:} For $l = 1, 2,\ldots, L_0+L_1-m-1$, set
\begin{eqnarray*}
\bs_l^{(m+1)} = \cases{ \bs_l^{(m)} &\quad
\mbox{if } $l< l^{(m)}$,\vspace*{2pt}
\cr
\bs_l^{(m)}\cup
\bs_{l+1}^{(m)} &\quad\mbox{if } $l=l^{(m)}$,\vspace*{2pt}
\cr
\bs_{l+1}^{(m)} &\quad\mbox{if } $l>l^{(m)}$. }
\end{eqnarray*}
\item[]{\bf Step 3:} Set $m = m+1$.
\end{itemize}
\item[]{\bf Output:} $\{\bs_l^{(m)}\}_{l=1}^{L_0+L_1-m}$ for $m = 1, 2,
\ldots, L_0+L_1-2$.
\end{itemize}

\begin{figure}

\includegraphics{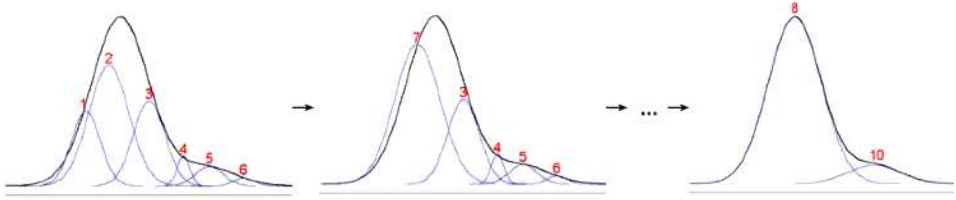}

\caption{An illustration of the HODC algorithm for six
density components: the HODC starts with clustering densities 1 and 2
as a mixture density labeled as 7, since the ``distance'' between 1 and
2 is shorter than all other adjacent density pairs. Then the HODC
computes the ``distance'' between densities 3 and 7, densities 3 and 4,
\ldots, to proceed the clustering. Following this procedure, the HODC
ends up with clustering densities 1, 2, 3 as a mixture density (labeled
as 8) and 4, 5, 6 as another mixture density (labeled as 10).}\label{fig:HODC}
\end{figure}

Figure~\ref{fig:HODC} illustrates the HODC algorithm. The algorithm
stops when $m = L_0+L_1-2$, where the ordered density components are
partitioned into two classes indexed by $\bs^{(m)}_{1}$ and $\bs
^{(m)}_{2}$. This\vspace*{-1pt} suggests that the number of indices in $\bs
_{k+1}^{(m)}$, denoted by $|\bs_{k+1}^{(m)}|$, is an estimate for $L_k$
in model \eqref{eq:finiteModel}. By running the HODC, we can obtain one
$L_k$ estimate for each posterior sample generated from a DPM fitting.
We take the average of $L_k$ estimates over all the posterior samples
as the input of NET-DPM-2. The HODC also provides an approximation to
$f_k(r)$ in \eqref{eq:f_k}, that is, $\wphi(r;\bs_{k+1}^{(m)},\cP)$.
This implies that we can further simplify the computation with the
algorithm in the following section.
%

\subsubsection{FMM guided by a DPM model fitting}
From a DPM model fitting, we obtain $V$ posterior samples of the
parameters for the marginal density of $\br$. We denote them as $\cP_v
= \{(\hmu_{vg}, \hsigma^2_{vg}, \hp_{vg})\}_{g=1}^{L_{v0}+L_{v1}}$, for
$v = 1,2,\ldots, V$. For\vspace*{1pt} each $\cP_v$, the HODC algorithm partitions
$L_{v0}+L_{v1}$ components into two classes, where the class-specific
components are indexed by $\ba_{v,0}$ and $\ba_{v,1}$. This leads to
$V$ approximations of $f_k(r)$, that is, $\wphi(r; \ba_{v,k},\cP_v)$.
Given $f_k(r)$, our proposed gene selection model reduces to
%
\begin{equation}
[ r_i \mid z_i = k ] \stackrel{\mathrm{i.i.d.}} {\sim}
f_k(r)\label
{eq:simpleModel}
\end{equation}
for $i = 1,2,\ldots, n$ and $k = 0, 1$, and $\bz$ follows \eqref
{eq:zPrior}. To make inference on the posterior distribution of $\bz$
by combining all $V$ approximations of $f_k(r)$, we consider
%
\begin{equation}
\pi(\bz\mid\br) \approx\frac{1}{V} \sum_{v=1}^V
\pi (\bz \mid\br , \wbphi_v ),
\end{equation}
where $\wbphi_v = \{\wphi(r; \ba_{v,0},\cP_v), \wphi(r; \ba
_{v,1},\cP
_v)\}$. For each $v$, the full conditional of $z_i$ is given by
%
\begin{eqnarray}\label{eq:fast_z_cond}
&&\pi(z_i = k \mid\bz_{-i}, \br,
\wbphi_v)  \nonumber
\\[-8pt]\\[-8pt]
&&\quad\quad\propto\wphi(r_i; \ba_{v,k},\cP_v)\exp
\biggl( \widetilde\omega _i \log (\pi_{k}) +
\varrho_{k} \sum_{j \neq i}
\omega_j c_{ij} I[z_j = k] \biggr).
\nonumber
\end{eqnarray}
We refer to this algorithm as NET-DPM-3 (see Section B.3 in the
supplemental article [\citet{zhao2014bayesian}] for details). It is
extremely fast with a moderate~$V$. Since the marginal density is
estimated without using the network information, it might introduce
bias on the distribution of $z_i$ and underestimate the variability of
$z_i$. From our experience, those issues do not affect the selection
accuracy much. Some examples are provided in Section~\ref{sec: simulation}.

\subsection{The choice of hyperparameters}
To proceed NET-DPMs, we need to specify the hyperparameters $\bpi$,
$\bvarrho$ and $\bomega$ in \eqref{eq:zPrior}. We assume that
$\bomega$
is prespecified according to biological information. In this paper, we
choose equal weight, that is, $\bomega= \bone_n$ without incorporating
any biological prior knowledge. We suggest two approaches to choosing
$\bpi$ and $\bvarrho$: (1) we assign hyperpriors on $\bpi$ and
$\bvarrho
$ and make posterior inference; (2) for a set of possible choices of
$\bpi$ and $\bvarrho$, we employ the Bayesian model averaging. The
details are provided in Section C in the supplemental article [\citet
{zhao2014bayesian}].

\section{Application}\label{sec:app}
To demonstrate the behavior of our method, we apply the proposed method
to the analysis of the Spellman yeast cell cycle microarray data set
[\citet{spellman1998comprehensive}]. The data set is intended to detect
genes with periodic behavior along the procession of the cell cycle. It
has been extensively used in the development of computational methods.
The network is summarized from the Database of Interacting Proteins
(DIP) [\citet{xenarios2002dip}]. We use the high-confidence connections
between yeast proteins from the DIP. Eventually, the network contains
2031 genes, where the mean, median, maximum and minimum edges per gene
are 3.948, 2, 57 and 1, respectively.

There is no outcome variable in the cell-cycle data set. In this
demonstration we focus on the selection of genes with periodic
behavior
in light of the network. It is\vadjust{\goodbreak} known that such genes show different
phase shifts along the cell cycle and may not be correlated with each
other [\citet{yu2010exploratory}]. We first perform the Fisher's exact $G$
test for periodicity [\citet{wichert2004identifying}] for each gene. We
then transform the $p$-values to normal quantiles, $r_i=-\Phi^{-1} (p_i)$
for gene $i$. We apply the fully Bayesian inference (NET-DPM-1), one
fast computation approach (NET-DPM-3) and the standard DPM model
fitting (STD-DPM) to this data set. For the NET-DPM-1, set $\tau_0=10$,
$\tau_1=2$; following the results by STD-DPM, set $\gamma_k=\omu_k,
\xi
_k^2=\osigma_k^2, \beta_k=10, \alpha_k=\osigma^2_k/\xi_k^2+1$ with
$k=0,1$, where $\{ \omu_k\}$ and $\{\osigma^2_k\}$ are preliminary
estimations by the STD-DPM. We also conduct a sensitivity analysis for
the hyperparameters specification (see the details in Section E in the
supplemental article [\citet{zhao2014bayesian}]) to verify the
robustness of the proposed methods. For both methods, the choices of
$\pi_0$ and $\bvarrho$ for the model averaging algorithm are
$(0.75,0.8,0.85,0.9)$ and $(0.5,1,5,10,15)\times(0.5,1,5,10,15)$ with
restriction $\varrho_0<\varrho_1$. We run all the algorithms 5000
iterations with 2000 burn-in. In this article, the standard DPM
fitting is obtained by an R package: {\tt{DPpackage}} and all the
proposed algorithms are implemented in R.




Table~\ref{table:real} presents the gene selection results based on
three methods in a two-by-two table format. The number of the
``selected'' genes by the NET-DPM-1, the NET-DPM-3 and the STD-DPM are
201, 216 and 114, respectively. The summation of the diagonal elements
of the table comparing the NET-DPM-3 and the NET-DPM-1 is larger than
that for NET-DPM-3 and the STD-DPM. This indicates a stronger agreement
between the two algorithms for NET-DPM.

\begin{table}
\tablewidth=260pt
\caption{Gene selection results by the three methods for the cell
cycle data set}\label{table:real}
\begin{tabular*}{260pt}{@{\extracolsep{\fill}}lcccc@{}}
\hline
& \multicolumn{2}{c}{\textbf{NET-DPM-1}}&\multicolumn
{2}{c@{}}{\textbf
{STD-DPM}} \\[-5pt]
& \multicolumn{2}{c}{\hrulefill}&\multicolumn{2}{c@{}}{\hrulefill}\\
& \textbf{Selected}& \textbf{Unselected}& \textbf{Selected}& \textbf
{Unselected}\\
\hline
NET-DPM-3 &&&& \\
\quad Selected &170 & \hphantom{00}46& 100 & \hphantom{0}116\\
\quad Unselected &\hphantom{0}31& 1784 & \hphantom{0}14 & 1801 \\
\hline
\end{tabular*}
\end{table}

\begin{figure}

\includegraphics{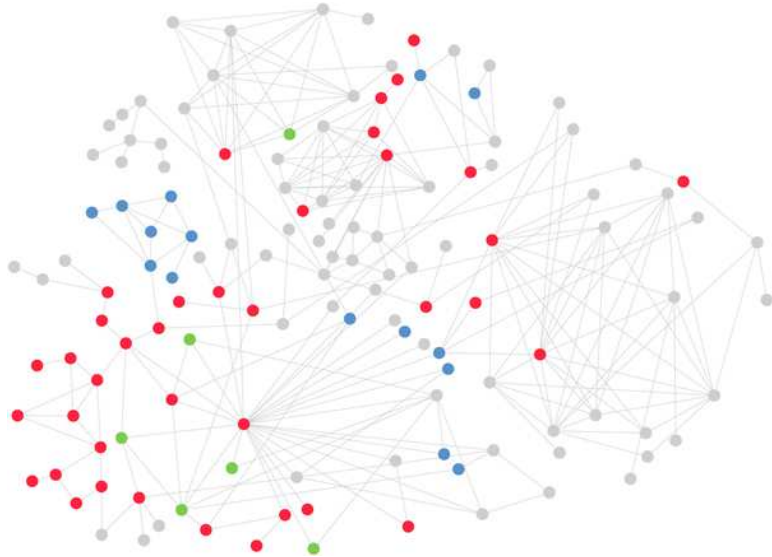}

\caption{A subnetwork composed of genes with periodic behavior. The subnetwork
consists of 135 genes. Red nodes: genes functionally involved in the
M-phase of cell cycle; blue node: genes functionally involved in the
interphase of cell cycle; green nodes: genes functionally involved in
both M and interphase of cell cycle.}\label{fig:app}
\end{figure}

We focus our discussion on the NET-DPM-3 results. After removing all
unselected genes, as well as selected genes not connected to any other
selected genes, 163 of the 216 genes fall into 11 subnetworks. Of the
11 subnetworks, 10 are very small, each containing 5 or less genes. The
remaining subnetwork contains 135 genes. Considering the purpose of the
study is to find genes with periodic behavior, and most such genes are
functionally related and regulated by the cell cycle process, this
result is expected. We present the subnetwork in Figure~\ref{fig:app}.
Sixty-one of the 135 genes belong to the mitotic cell cycle process
based on gene ontology [\citet{ashburner2000gene}]. The yeast mitotic
cell cycle can be roughly divided into the M phase and the interphase,
which contains S and G phases [\citet{ashburner2000gene}]. We do not
further divide the interphase because the number of genes annotated to
its descendant nodes are small. Among the 135 genes, 45 are annotated
to the M phase, and 21 are annotated to the interphase. By coloring the
M phase genes in red, the interphase genes in blue and the genes
annotated to both phases in green, we see that the majority of the
selected M phase genes are clustered on the subnetwork, while the
selected interphase genes are somewhat scattered, with 7 falling into a
small but tight cluster.

\begin{figure}[t!]

\includegraphics{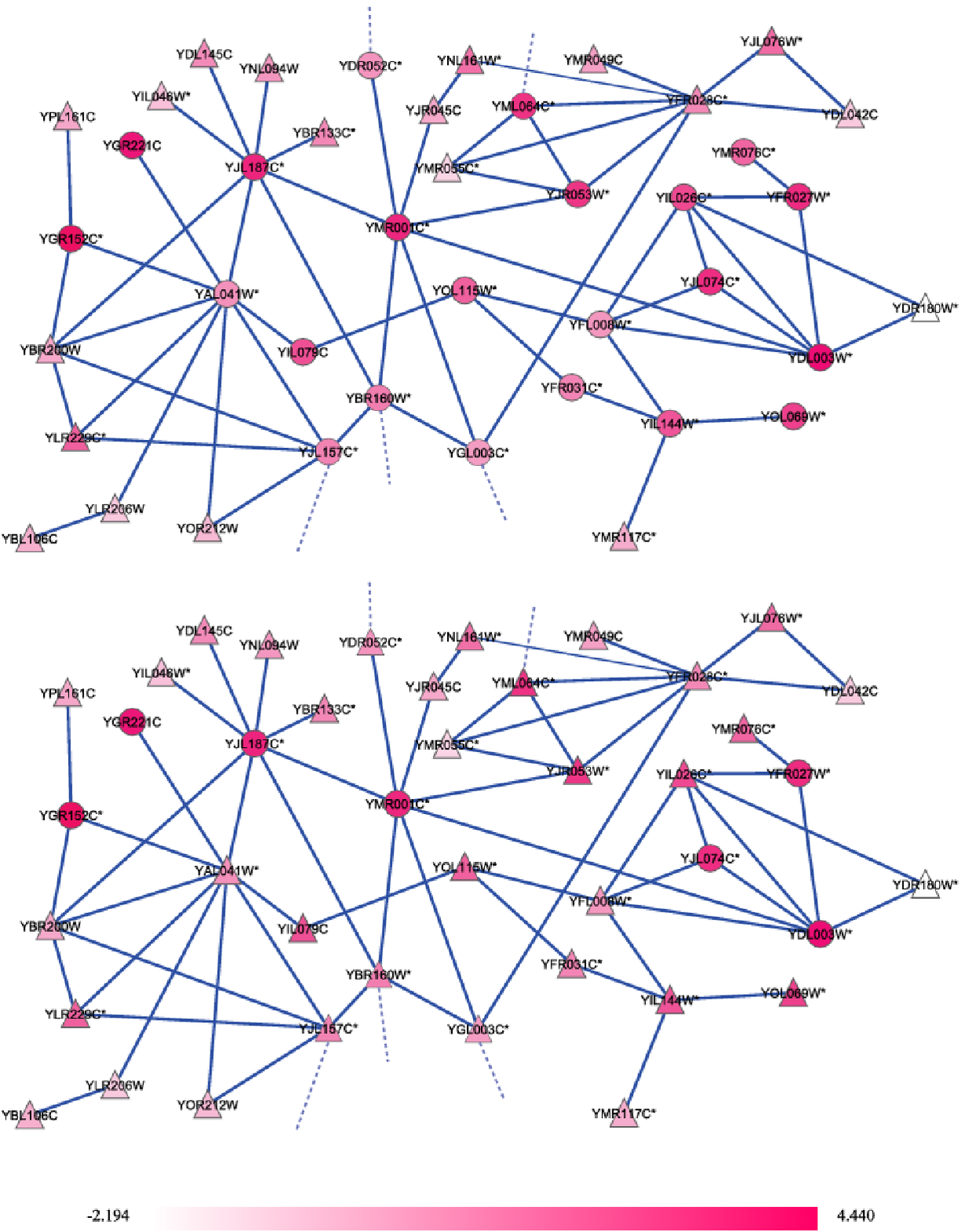}

\caption{A portion of the subnetwork shown in Figure \protect\ref
{fig:app},
together with the immediate neighbors of the selected genes.
Upper panel: NET-DPM-3 results; lower panel: STD-DPM results.
The node labels indicate the gene name; circles and triangles
represent
``selected'' and ``unselected'' genes; colors denote the value of the
normal
quantiles; a star in superscript represents the genes
functionally annotated to the cell cycle process. Dash
lines denote connections to genes not shown in the figure.}\label{fig:app2}
\end{figure}


We show part of the subnetwork detected by the NET-DPM-3 with the
corresponding one under the STD-DPM in Figure~\ref{fig:app2}, where the
genes that are linked by a dashed line are connected to other genes
that are not shown in the figure.
In this subnetwork, the gene selection results by the NET-DPM-1 agree
with the NET-DPM-3 except for only one gene ``YML064'' for which the
NET-DPM-1 does not select it with probability 0.478, while the
NET-DPM-3 selects it with probability 0.687. This implies that both
methods provide large uncertainty on this gene. Comparing the top panel
(our method, NET-DPM-3) and bottom panel (STD-DPM), we observe a number
of genes selected by NET-DPM but not by STD-DPM, and almost all such
genes are cell cycle-related (denoted by a star by the ORF name).
Examples include YAL041W (CLS4), which is required for the
establishment and maintenance of polarity and critical in bud formation
[\citeauthor{chenevert1994identification} (\citeyear{chenevert1994identification});
\citeauthor{cherry2012saccharomyces} (\citeyear{cherry2012saccharomyces})]. The gene
only shows moderate periodic behavior, as denoted by the color of the
node. However, due to its links to other genes that have strong
periodic behavior, it is selected by our method as an interesting gene.
Another example is YFL008W (SMC1). It is a subunit of the cohesion
complex, which is essential in sister chromatid cohesion of mitosis and
meiosis. The complex is also involved in double-strand DNA break repair
[\citeauthor{strunnikov1999structural} (\citeyear{strunnikov1999structural});
\citeauthor{cherry2012saccharomyces} (\citeyear{cherry2012saccharomyces})]. Similar to
CLS4, the periodic behavior of SMC1 is not strong enough. It is only
selected when the information is borrowed from linked genes that are
functionally related and show strong periodic behavior. A number of
other cell cycle-related genes in Figure~\ref{fig:app2} are in a
similar situation, for example, YBR106W, YDR052C, YJL157C, YGL003C and
YMR076C. These examples clearly show the benefit of utilizing the
biological information stored in the network structure.

To assess the functional relevance of the selected genes globally, we
resort to mapping the genes onto gene ontology biological processes
[\citet{ashburner2000gene}]. We limit our search to the GO Slim terms
using the mapper of the Saccharomyces Genome Database [\citet
{cherry2012saccharomyces}]. The full result is listed in the
supplementary file. Clearly, the overrepresented GO Slim terms are
centered around cell cycle. Here we discuss some GO terms that are
nonredundant. Among the 216 selected genes, 70 (32.4\%, compared to
4.5\% among all genes) belong to the process response to DNA damage
stimulus (GO:0006974). The term shares a large portion of its genes
with DNA recombination (GO:0006310) and DNA replication (GO:0006260)
processes, which are integral to the cell cycle. Sixty-seven of the
selected genes (31.0\%, compared to 4.7\% among all genes) belong to
the process mitotic cell cycle (GO:0000278). Twenty-six of the 67 genes
are shared with response to DNA damage stimulus (GO:0006974). Forty-one
of the selected genes (19.0\%, compared to 3.0\% among all genes)
belong to the process regulation of cell cycle (GO:0051726), among
which 29 also belong to mitotic cell cycle (GO:0000278). Thirty-one of
the selected genes (14.4\%, compared to 2.6\% among all genes) belong
to the process meiotic cell cycle (GO:0051321), among which 12 are
shared with mitotic cell cycle (GO:0000278). Other major enriched terms
include chromatin organization (12.5\%, compared to 3.5\% overall),
cytoskeleton organization (12.5\%, compared to 3.4\% overall),
regulation of organelle organization (9.7\%, compared to 2.4\% overall)
and cytokinesis (7.9\%, compared to 1.7\% overall). These terms clearly
show strong relations with the yeast cell cycle.

\section{Simulation studies}\label{sec: simulation}
In this section we illustrate the performance of our methods (NET-DPMs)
using simulation studies with various network structures and data
settings compared with other methods. In Simulation\vadjust{\goodbreak} 1 we study the
similarity between the fully computational algorithm NET-DPM-1 and two
fast computation approaches NET-DPM-$x$, $x=2,3$, in terms of gene
selection accuracy and uncertainty estimations. Each of the three
algorithms can be used along with one of the two methods for choosing
hyperparameters: the posterior inference and model averaging. In
Simulation 2 we focus on the gene network selection under a particular
network structure and two types of simulated data to demonstrate the
flexibility of the proposed methods. In both simulations we compare the
NET-DPMs with a STD-DPM combined with the HODC algorithm without using
any network information. In Section D in the supplemental
article [\citet{zhao2014bayesian}], we also demonstrate the flexibility of the
proposed methods by conducting a simulation on the selection of genes
that are strongly associated with an outcome variable, and compare our
NET-DPMs with a network based Bayesian variable selection (NET-BVS)
proposed by \citet{li2010bayesian}.
%
\begin{figure}[b]

\includegraphics{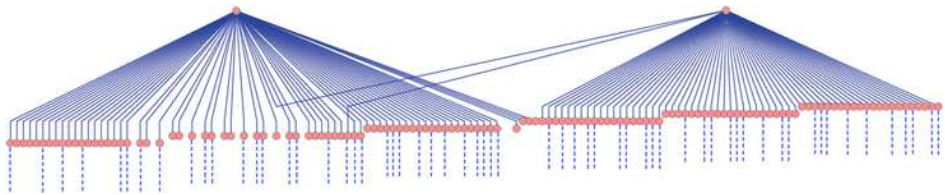}

\caption{Partial network structure with the
dash lines representing connections to other nodes not shown in the
figure.}\label{figure:simu1_sub}
\end{figure}

\subsection{Simulation 1}\label{subsec:s1}

\begin{figure}

\includegraphics{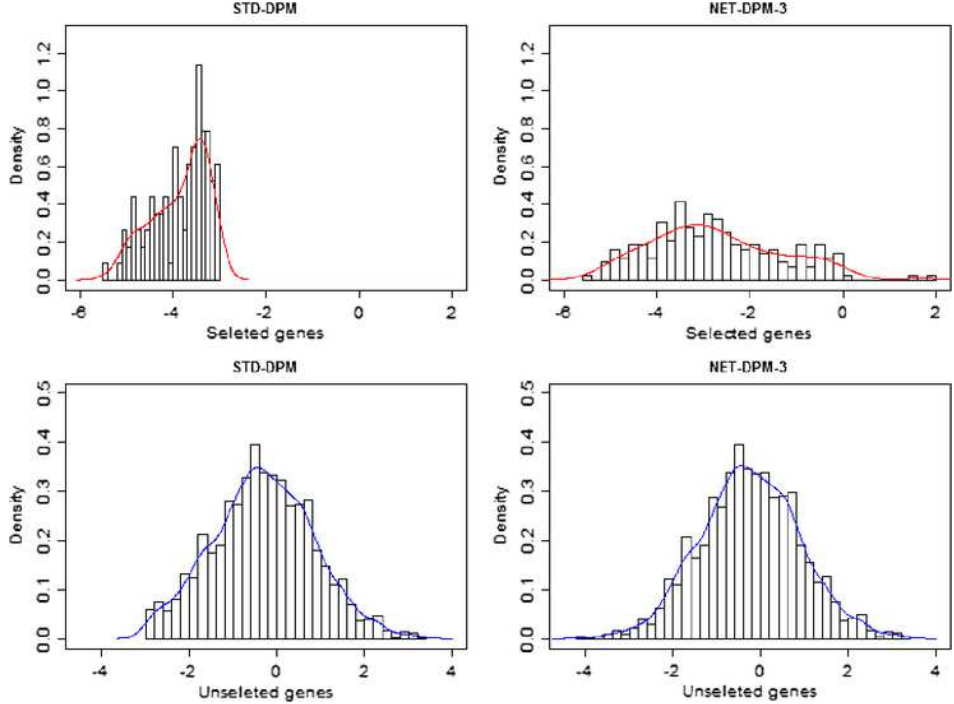}

\caption{Empirical distributions of ``selected''
genes (upper panel) and ``unselected'' genes (lower panel) in the
Spellman yeast cell cycle data estimated by the NET-DPM-3 (right panel)
and the STD-DPM (left panel).}\label{fig:appdensity}
\end{figure}

In this simulation we investigate the performance of the proposed
algorithms using a simulated data set that mimic the real data in
Section~\ref{sec:app}. We generate a scale-free network with 1000
genes based on the rich-get-rich algorithm [\citet
{barabasi1999emergence}], that is, $n = 1000$. Two hub genes with 64
and 69 connections to other genes are in this network; the mean and
median edges per gene are 1.998 and 1. The partial network structure
with the two hub genes included is shown in Figure~\ref{figure:simu1_sub}. From the network structure, we generate $\bz$ from
the Ising model \eqref{eq:zPrior} with the sparsity parameter $\pi
_0=0.8$ and smoothness parameters $\bvarrho=(\varrho_0, \varrho_1)=(5,
10)$. For $i = 1,\ldots, n$, in light of the results in Section~\ref{sec:app}, we simulate data $r_i$ given $z_i$ from the empirical
distributions (Figure~\ref{fig:appdensity}) of the test statistics for
``selected'' and ``unselected'' genes in the Spellman yeast cell cycle
microarray data. As shown in Section~\ref{sec:app}, the NET-DPM-3
(Scenario 1) and the STD-DPM (Scenario 2) provide different gene
selections results. We set both scenarios as the truth to simulate data.

\begin{figure}[b]

\includegraphics{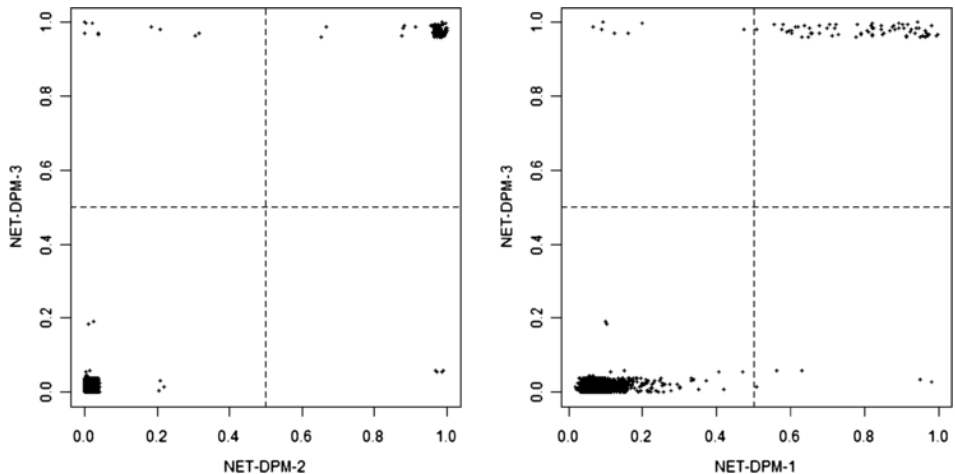}

\caption{Marginal posterior probabilities of the
class labels of all 1000 genes by the different methods: NET-DPM-3 vs.
NET-DPM-2 (left panel) and NET-DPM-2 vs. NET-DPM-1 (right panel). The
probability values are jittered by tiny random noises for better presenting.}\label{figure:simu1}
\end{figure}

We apply the NET-DPM-$x$, for $x=1,2,3$, and the STD-DPM to the
simulated data set. To choose the sparsity and smoothness parameters,
the NET-DPM-1 and the NET-DPM-3 are both combined with model averaging,
where the possible choices of $\pi_0$ and $\bvarrho$ are
$(0.75,0.8,0.85,0.9)$ and $(1,5,10,20,50)\times(1,5,10,20,50)$, while
the NET-DPM-2 is combined with the posterior inference on $(\pi
_0,\bvarrho)$. As for other hyperparameters, we specify $\tau_k, \xi_k,
\gamma_k, \beta_k, \alpha_k; k=0,1$ the same way as in the data
application for the NET-DPM-$x$, for $x=1,2$. With random starting
values, each algorithm is run 10 times under 10,000 iterations with
2000 burn-in. For each gene $i$, the mode of the marginal posterior
probability of $z_i$ is taken to determine whether gene $i$ is selected
or not. The selection performance for each method based on the average
of the 10 runs is presented in Table~\ref{table:simu1}. We also compare
the posterior probability estimates of $\bz$ between different
algorithms under Scenario 1 in Figure~\ref{figure:simu1}. 

\begin{table}
\tabcolsep=0pt
\caption{Gene selection accuracy in Simulation 1}\label{table:simu1}
\begin{tabular*}{\textwidth}{@{\extracolsep{\fill}}lcccc@{}}
\hline
& \textbf{STD-DPM} & \textbf{NET-DPM-1} & \textbf{NET-DPM-2} &\textbf{NET-DPM-3}\\
\hline
&\multicolumn{4}{c@{}}{Scenario 1}\\
True positive rate & 0.893 & 0.973 & 0.920 & 0.920\\
False positive rate & 0.292 & 0.001 & 0.000 & 0.006\\
False discovery rate & 0.801 & 0.014 & 0.000 & 0.080\\[3pt]
&\multicolumn{4}{c@{}}{Scenario 2}\\
True positive rate & 1.000 & 1.000 & 1.000 & 1.000\\
False positive rate & 0.232 & 0.000 & 0.000 & 0.007\\
False discovery rate & 0.741 & 0.000 & 0.000 & 0.085\\[6pt]
Typical computation time (hrs) & 0.100 & 8.500 & 2.800 & 0.150\\
\hline
\end{tabular*}
\end{table}

\begin{figure}

\includegraphics{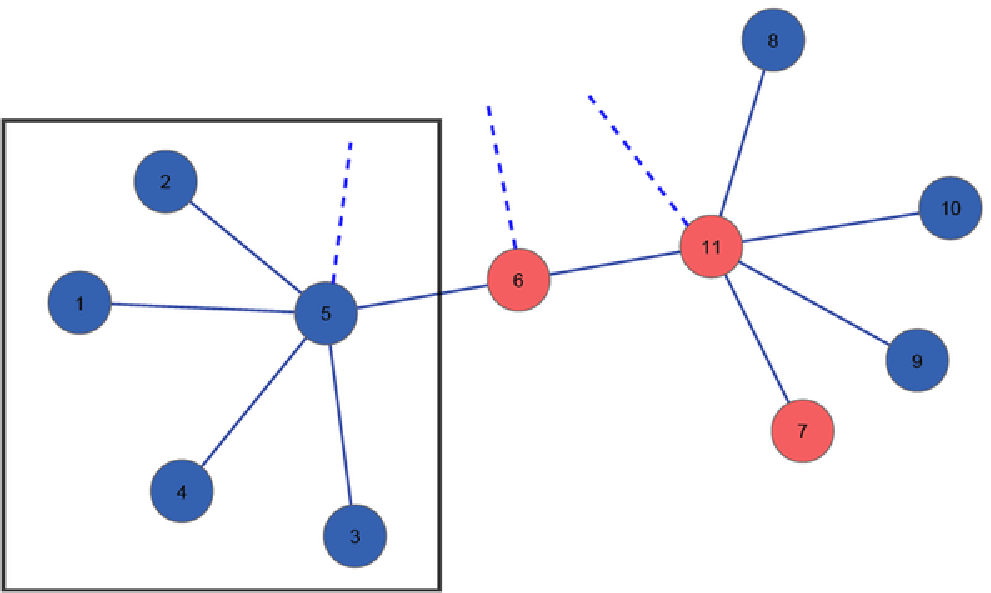}

\caption{Partial simulated gene network structure:
the blue nodes represent ``selected'' genes and the red nodes represent
``unselected'' genes. Dash lines denote connections to genes not shown
in the figure. A subnetwork of interest includes nodes 1, 2, 3, 4 and
5, which are encircled by a rectangle frame.}\label{figure:simu2}
\end{figure}

From Table~\ref{table:simu1}, it is clear that the NET-DPMs achieve a
better selection performance than the STD-DPM method under both
scenarios. The STD-DPM without using the gene network information
provides an extreme high false discovery rate in each scenario. This
implies that it is critical to incorporate the gene network information
to control FDR. Table~\ref{table:simu1} also suggests the NET-DPM-2 and
the NET-DPM-3 approximate the NET-DPM-1 very well in terms of the gene
selection accuracy with a substantial lower computational cost (3.4 GHz
CPU, 8~GB Memory,
Windows System). In addition, a comparison between the NET-DPM-2 and
the NET-DPM-3 shows that the Bayesian model averaging over
hyperparameters $(\pi_0,\bvarrho)$ provides an efficient alternative to
the standard Bayesian posterior inference procedure. For the posterior
probability estimates, the NET-DPM-2 and the NET-DPM-3 achieve a good
agreement, as shown in the left panel of Figure~\ref{figure:simu1}.
However, in the right panel of Figure~\ref{figure:simu1}, compared with
the NET-DPM-1, the NET-DPM-3 tends to provide larger probability
estimates for the ``selected'' genes, but smaller probability estimates
for ``unselected'' genes. This implies the fast computation approaches
underestimate the uncertainty of gene selection.

\subsection{Simulation 2}\label{subsec:s2}
In this simulation we demonstrate the flexibility of the proposed
methods and their ability to identify subnetworks of interest. We
consider a 94-gene network which consists of an 11-gene subnetwork by
design and an 83-gene scale-free network simulated from the
rich-get-rich algorithm. The mean and median edges per node for the
whole network are 2.02 and 1. Figure~\ref{figure:simu2} shows the
designed 11-gene subnetwork, where genes 5, 6 and 11 are connected with
three other genes from the 83 gene scale-free network. Rather than
simulating from priors, we directly specify the class label $\bz$ as
$z_i = 1$ for $i\in\{1,2,3,4, 5, 8,9,10\}$, $z_i = 0$ otherwise. In
Figure~\ref{figure:simu2} the blue nodes represent the ``selected''
genes and the red nodes are ``unselected'' genes. In addition, all other
genes in the scale-free network (not shown in the figure) are
``unselected.'' The gene subnetwork of interest includes genes 1, 2, 3,
4 and 5, which are encircled by a rectangle frame in Figure~\ref{figure:simu2}. The null distribution for ``unselected'' $r_i$ is
specified as a standard normal distribution: $ [ r_i \mid z_i = 0
]\sim\mN(0, 1)$. For the distribution of ``selected'' genes, we
consider two settings:
\begin{eqnarray*}
\mbox{Gaussian data:}&\quad\quad& [ r_i \mid z_i = 1 ] \sim0.4
\times\mN (3,1)+0.6\times\mN(2,0.5),
\\
\mbox{Non-Gausian data:}&\quad\quad& [ r_i \mid z_i = 1 ]
\sim0.4\times \mG (5,2)+0.6\times\mG(6,3),
\end{eqnarray*}
where $\mG(a,b)$ denotes a gamma distribution with shape $a$ and rate
$b$. According to the above procedure, we simulate 100 data sets for
each type of data. We apply the NET-DPM-3 and the STD-DPM to each data
set. We utilize the model averaging for choosing hyperparameters and a
set of possible choices are given by $\{1, 2, 5, 10, 15\}$ for both
$\varrho_0$ and $\varrho_1$, and \{0.8, 0.85, 0.9, 0.95\} for $\pi_0$.
We run 10,000 iterations with 2000 burn-in on each data set for both
methods. In each simulated data set, we predetermine one gene as a
``sure selected'' gene. It has the largest number of connections with
the ``selected'' genes estimated by the STD-DPM model.

Table~\ref{table:simu2} summarizes the selection accuracy of the gene
subnetwork based on the 100 simulated data sets for each type of data.
It is clear that the NET-DPM-3 provides much higher accuracy of the
subnetwork selection than the STD-DPM. The NET-DPM-3 achieves a more
than 60\% accuracy rate in correctly identifying the subnetwork with an
additional low false positive and false negative occurrences regardless
of the type of data. This verifies the overall better performance of
NET-DPM-3 than the STD-DPM in terms of identifying the gene subnetwork
and the robustness of the proposed methods on different types of data.


\begin{table}
\caption{Selection accuracy of gene subnetwork\protect\tabnoteref{ta}
by TPR (true positive rate), FPR (false positive rate)
and FDR (false discovery rate) in Simulation 2}\label{table:simu2}
\begin{tabular*}{\textwidth}{@{\extracolsep{\fill}}lcccccc@{}}
\hline
& \multicolumn{3}{c}{\textbf{Gaussian data}} & \multicolumn
{3}{c@{}}{\textbf{Non-Gaussian data}}\\[-5pt]
& \multicolumn{3}{c}{\hrulefill} & \multicolumn{3}{c@{}}{\hrulefill
}\\
\textbf{Method}&\textbf{TPR}& \textbf{FPR}& \textbf{FDR} & \textbf{TPR}& \textbf{FPR} & \textbf{FDR}\\
\hline
NET-DPM-3 & 63\%& 11\% & 15\% & 60\%& \hphantom{0}5\% &\hphantom{0}8\%
\\
STD-DPM & 15\% & 33\% & 69\% & 17\% & 26\% & 60\%\\
\hline
\end{tabular*}
\tabnotetext{ta}{For gene subnetwork selection, the TPR is defined as
the percentage of exactly selecting the correct network. The FPR is the
percentage of selecting a larger network containing the correct network
and at least one more other gene that has connection to the network.
The FDR is the proportion of falsely selecting a larger network among
all the network discoveries (selecting a correct or larger network).}
\vspace*{-3pt}
%
%
\end{table}

\section{Discussion}\label{sec:disc}

In the article we propose a Bayesian nonparametric mixture model for
gene/gene subnetwork selection. Our model extends the standard DPM
model incorporating the gene network information to significantly
improve the accuracy of the gene selections and reduce the false
discovery rate. We demonstrate that the proposed method has the ability
to identify the subnetworks of genes and individual genes with a
particular expressional behavior. We also show that it is able to
select genes which are strongly associated with clinical variables. We
develop a posterior computation algorithm along with two fast
approximation approaches. The posterior inference can produce more
accurate uncertainty estimates of gene selection, while the fast
computing algorithms can achieve a similar gene selection accuracy. Due
to the nonparametric nature, our method has the flexibility to fit
various data types and has robustness to model
assumptions.\vadjust{\goodbreak}

When we observe gene expression data along with measurements of a
clinical outcome, we need to create statistics to perform the selection
of genes that are strongly associated with the clinical outcome. The
choice of the statistics is crucial to the performance of our methods.
To model the relationship between the clinical outcome and gene
expression data, much literature suggests a linear regression model
[\citeauthor{li2008network} (\citeyear{li2008network});
\citeauthor{pan2010incorporating} (\citeyear{pan2010incorporating});
\citeauthor{li2010bayesian} (\citeyear{li2010bayesian});
\citeauthor{stingo2011incorporating} (\citeyear{stingo2011incorporating})], from which we produce testing statistics or
coefficient estimates as the candidates. For instance, as we suggest in
Section D in the supplemental article [\citet{zhao2014bayesian}], the
most straightforward approach is to fit simple linear regression on
each gene and use the $t$-statistics as the input data to our methods.
However, there is no scientific evidence that the relationship between
gene expression profiles and the clinical outcome should follow a
linear regression model. Without making this assumption, we may test
the independence between each gene expression profile and the clinical
outcome via a nonparametric model suggested by \citet{einmahl2008tests}
and use our model to fit the testing statistics. Other potential
choices of statistics for the nonlinear problems include mutual
information statistics [\citet{peng2005feature}] and maximal information
coefficient (MIC) statistics [\citet{reshef2011detecting}].

Although the development of our method is motivated by gene selection
problems, our method can conduct variable selection for a general
purpose and it has broad applications. For example, functional
neuroimaging studies (e.g., fMRI and PET) usually produce large-scale
statistics, one for each voxel in the brain. Those statistics are used
to localize the brain activity regions related to particular brain
functions. This essentially is a voxel selection problem to which our
method is applicable, where the networks may be defined according to
the spatial locations of the voxels. In addition to this, we discuss
two future directions:
\begin{longlist}
\item[(1)] It is common that we have multiple
hypothesis tests for each gene, and we have interest in jointly
analyzing these statistics. This motivates an extension of the current
NET-DPM model from one dimension to multiple dimensions for
multivariate large-scale statistics. \item[(2)] The selection of one gene
might be affected by not only the genes that are directly connected to
it, but also the genes close to it over the network. It would be
interesting to extend the prior specifications of the class label by
incorporating a network distance. This should provide more biologically
meaningful results.
\end{longlist}

\section*{Acknowledgments}
The authors would like to thank the Editor, the
Associate Editor and two referees for their
helpful suggestions and constructive comments
that substantially improved this manuscript.

\begin{supplement}[id=suppA]
\stitle{Supplement to ``A Bayesian nonparametric mixture model for
selecting genes and gene subnetworks''}
\slink[doi]{10.1214/14-AOAS719SUPP} 
\sdatatype{.pdf}
\sfilename{aoas719\_supp.pdf}
\sdescription{In this online supplemental article we provide (A)
derivations of the proposed methods, (B) details of the main algorithms
for posterior computations, (C) details of posterior inference for
hyperparameters, (D) additional simulation studies and (E) sensitivity
analysis.}
\end{supplement}

%

\printaddresses

\end{document}